\documentclass[twocolumn,usenatbib]{aastex62}
\usepackage{subfigure}
\usepackage{longtable}
\usepackage{graphicx}
\usepackage{graphics}
\usepackage{keyval}
\usepackage{trig}
\usepackage[american]{babel}
\usepackage{url}
\usepackage{color}
\usepackage{amsmath}
\usepackage{bm}

\def\beq{\begin{equation}}
\def\eeq{\end{equation}}
\def\bey{\begin{eqnarray}}
\def\eey{\end{eqnarray}}

\def\msun{M_\odot}
\def\sun{\odot}
\def\lsim{\mathrel{\raise.3ex\hbox{$<$\kern-.75em\lower1ex\hbox{$\sim$}}}}
\def\gsim{\mathrel{\raise.3ex\hbox{$  $\kern-.75em\lower1ex\hbox{$\sim$}}}}

\def\tz{t_\mathrm{0}}
\def\te{t_\mathrm{E}}

\def\uz{u_\mathrm{0}}

\def\thetae{\theta_\mathrm{E}}

\def\ds{D_\mathrm{S}}
\def\dl{D_\mathrm{L}}

\def\a0{A_{\mathrm{0}}}

\def\sout{s_\mathrm{out}}

\def\arcsec{^{\prime\prime}}
\def\deltapar{\delta_{\parallel}}
\def\deltaperp{\delta_{\perp}}

\def\c2dof{\chi^2/{\rm d.o.f.}}

\def\m606{m_{\rm F606W}}
\def\m814{m_{\rm F814W}}

\def\nd{N_{\rm D}}
\def\np{N_{\rm P}}

\def\dbic{\Delta_\mathrm{BIC}}

\def\apjl{ApJ}
\def\aap{A\&A}      % Astronomy and Astrophysics (Letters indicated by number)
\def\apjs{ApJS}       % Astrophysical Journal Supplement Series (the)
\newcommand\Eq[1]{Eq.~(\ref{#1})}
\newcommand\Fig[1]{Fig.~\ref{#1}}
\newcommand\Tab[1]{Table~\ref{#1}}
\newcommand\Sec[1]{Sec.~\ref{#1}}
\begin{document}

\title{A search for black hole microlensing signatures in globular cluster NGC~6656 (M~22)}

\author{N. Kains}\affil{Space Telescope Science Institute, 3700 San Martin Drive, Baltimore, MD 21218}\affil{Department of Physics \& Astronomy, Barnard College, Columbia University, 3009 Broadway, New York, NY 10027}
\author{A. Calamida}\affil{Space Telescope Science Institute, 3700 San Martin Drive, Baltimore, MD 21218}
\author{K. C. Sahu}\affil{Space Telescope Science Institute, 3700 San Martin Drive, Baltimore, MD 21218} 
\author{J. Anderson}\affil{Space Telescope Science Institute, 3700 San Martin Drive, Baltimore, MD 21218}  
\author{S. Casertano}\affil{Space Telescope Science Institute, 3700 San Martin Drive, Baltimore, MD 21218}  
\author{D.M. Bramich}\affil{New York University Abu Dhabi, PO Box 129188, Saadiyat Island, Abu Dhabi, UAE}

\begin{abstract}
We report the results of a study aiming to detect signs of astrometric microlensing caused by an intermediate-mass black hole (IMBH) in the center of globular cluster M~22 (NGC~6656). We used archival data from the \textit{Hubble Space Telescope (HST)} taken between 1995 and 2014, to derive long-baseline astrometric time series for stars near the center of the cluster, using state-of-the-art software to extract high-precision astrometry from images. We then modelled these time-series data, and compared microlensing model fits to simple linear proper-motion fits for each selected star. We find no evidence for astrometric microlensing in M~22, in particular for Bulge stars, which are much more likely to be lensed than cluster stars, due to the geometry of microlensing events. Although it is in principle possible to derive mass limits from such non-detections, we find that no useful mass limits can be derived for M~22 with available data, mostly due to a 10-year gap in coverage. This is a result from difficulties with deriving precise enough astrometry from Wide Field Planetary Camera 2 (WFPC2) observations, for stars that do not fall on the PC chip. However, this study shows that, for other \textit{HST} instruments, we are able to reach precisions at which astrometric microlensing signals caused by IMBH in globular clusters could be detected, and that this technique is a promising tool to make a first unambiguous detection of an IMBH.

\end{abstract}

\keywords{microlensing --- astrometry --- black holes --- exoplanets --- surveys}

% ====================================================
\section{Introduction} \label{sec:intro}
% ====================================================

The formation of supermassive black holes (SMBHs) in the early Universe is one of the most debated issues in modern astronomy. Indeed, SMBHs are found at large redshifts, meaning that some of them had already formed only a few hundred million years after the Big Bang \citep{fan06}. However, this quick formation cannot be explained by accretion alone, since stellar-mass black holes cannot accrete enough matter in such a short time, even at the maximum allowed accretion rate, the Eddington rate. Although super-Eddington accretion might take place via different mechanisms \citep[e.g.][]{alexander14}, the favoured scenario for SMBH formation is instead through the merger of seed intermediate-mass black holes (IMBHs), which themselves may form through runaway mergers of stars.

For this reason, much effort has been devoted in recent years to searching for evidence of the existence of IMBHs, focusing on environments where they might have formed, such as in
globular clusters. Besides having high enough stellar densities to form IMBHs, globular clusters have velocity dispersions consistent with a low-mass extrapolation of the M-$\sigma$ relation \citep[e.g.][]{ferrarese00}, making them good host candidates for IMBHs. In addition to this, they could also have delivered seed IMBHs to the center of galaxies to form SMBHs, since globular clusters are usually about the same age as their host galaxies.

\cite{kains16} proposed using gravitational microlensing to achieve an unambiguous IMBH detection, motivated by hitherto-published detections being subject to strong caveats and possible alternative explanations. Indeed, microlensing by an IMBH in the cluster would produce an astrometric deflection to the position of background stars, which, if detected, would then allow us to derive a mass for the IMBH, without relying on any assumption on the nature of the system. This is an advantage over other techniques; for instance, observations aiming to detect X-ray emission caused by accretion onto an IMBH are reliant on assumed accretion models \citep{grindlay01, maccarone05, haggard13, mezcua15}, and cusps in both photometric and kinematic profiles are not unique signature of the presence of an IMBH in the cluster center \citep{illingworth77, baumgardt03, lanzoni07, ibata09, trenti10, vesperini10}.

In this paper, we report attempts to detect the astrometric deflection that might be caused by an IMBH in the globular cluster M~22 on background stars, due to gravitational microlensing. M~22 (NGC~6656) was identified by \cite{kains16} as the most promising candidate for such a detection, thanks to the combination of a high density of background Galactic Bulge stars, a large cluster-Bulge relative motion, and, at 3.2 kpc, being one of the closest globular clusters in the Milky Way. This makes the detection of an astrometric lensing event more likely to be achievable over reasonable timescales, given their long typical timescales ($\sim3000$ days for an $10^4\msun$ IMBH).

In \Sec{sec:microlensing}, we review the basics of astrometric microlensing, and how a detection of this effect can lead to a direct mass estimate; in \Sec{sec:data}, we describe the \textit{HST} archival data used for this study, and the reduction process. We outline our procedure to model astrometric time series for each star in our sample in \Sec{sec:modelling}, and discuss our results in \Sec{sec:results}.

\section{Astrometric microlensing}\label{sec:microlensing}

For a full discussion of astrometric microlensing, we refer the reader to \cite{hog95} and \cite{dominik00}. Here we recall the key equations that are relevant to the work presented in this paper.

Gravitational microlensing occurs when a background source, located at a distance $\ds$ from the observer, becomes sufficiently aligned with a foreground \textit{lens} object at distance $\dl$. Given a lens of mass $M$, the Einstein ring radius $\thetae$ gives a typical angular scale for the alignment necessary for microlensing to occur, and is given as

\begin{equation}\label{eq:thetae}
\thetae = \sqrt{\frac{4GM}{c^2}(\dl^{-1} - \ds^{-1})}\, .
\end{equation}

Because multiple, non-identical, images of the source are produced during a microlensing event, but these usually cannot be resolved, we can, instead of the motion of individual images, observe the motion of the source's
centroid as the event unfolds. With the source-lens angular separation $u$ expressed in units of $\thetae$, one can derive an expression for the astrometric shift $\delta(u)$ that occurs during a microlensing event \citep{hog95} as 

\begin{equation}\label{eq:deltau}
\delta(u)=\frac{u}{u^2+2}\thetae\, .
\end{equation}

Note that this expression assumes a dark lens, i.e. no or negligible blended light from the lens. This gives the total deflection caused by microlensing, which points away from the lens, as seen by the observer, i.e. along a line joining the lens and undeflected source positions.

The components parallel and perpendicular to the source-lens relative motion can be expressed (e.g. \citealt{dominik00}) as

\begin{gather}
\deltapar=\frac{p}{\uz^2 + p^2 + 2}\thetae \nonumber \\
\deltaperp=\frac{\uz}{\uz^2 + p^2 + 2}\thetae \label{eq:deltacomp}\, ,
\end{gather}

\noindent
where $\uz$ is the impact parameter, or minimum source-lens angular separation, in units of $\thetae$, and $p \equiv p(t)$ is expressed as 

\begin{equation}\label{eq:p}
p=\frac{t-\tz}{\te}\, ,
\end{equation}

\noindent
where $\tz$ is the time at which $u=\uz$, $\te$ is the Einstein timescale, defined as the time taken by the source to cross $\thetae$. In two dimensions, a source affected by microlensing traces an elliptical motion, in the source's rest frame, and relative to its undeflected position, with eccentricity $\epsilon=[2/(u_0^2 + 2)]^{1/2}$ \citep{dominik00}.

From \Eq{eq:thetae}, we see that the mass of the lens, $M$, can be measured directly if three quantities are determined: the distances to the source and lens, $\ds$ and $\dl$, and $\thetae$. Although the lens distance is usually unknown in microlensing events, and can only be constrained when second-order effects can be measured from photometric light curves, such as the parallax effect caused by the Earth's orbit around the Sun \citep[e.g.][]{dominik98, an02, gould04}, in the case of an IMBH in the center of a globular cluster, $\dl$ is known, to the extent that the distance to the cluster is known. For Galactic globular clusters, this is usually the case to a precision of a few hundred parsecs \citep[e.g.][]{harris96}.

The discussion above assumes that the images are unresolved. However, when considering IMBH lenses, with possible values of $\thetae$ of several hundreds of mas for the top end of the IMBH mass range ($> 10^5 \msun$), it is possible that some events may fall in the ``partially-resolved'' regime. In this case, the separation between the source images is large enough to be resolved with instruments on \textit{HST}, and \Eq{eq:deltau} becomes instead

\begin{equation}\label{eq:deltaupr}
\delta(u)_{pr}=\frac{1}{2}(\sqrt{u^2+4}-u)\thetae\, .
\end{equation}

For values of $u\gtrsim4$, Equations \ref{eq:deltau} and \ref{eq:deltaupr} converge. In the following discussion of our data and modelling procedure, we will show that here we can safely assume that stars in the unresolved microlensing regime.

\section{Data and reduction process}\label{sec:data}

\subsection{Archival data}

We searched the Mikulski Archive for Space Telescopes (MAST) for data that would satisfy several criteria. Images must have been observed in medium- or wide-band filters, with the Wide Field Planetary Camera 2 (WFPC2), the Advanced Camera for Survey (ACS)'s, or Wide Field Camera 3 (WFC3); this is because observations taken with narrow-band filters do not go deep enough to enable us to derive precise astrometry for Bulge stars. Images must also contain the center of M~22 at least 10$\arcsec$ from the edge of the image, and have exposure times long enough to probe background Bulge stars located behind the cluster. A summary of the final data set satisfying these criteria, spanning 19.33 years, is given in \Tab{tab:data}.

%% =====================================================
\begin{table*}
  \vspace{0.5cm}
\begin{center}
  \begin{tabular}{cccccccc}
\hline
Proposal ID  	&PI	&Instrument/ Camera	&Dates	&Filters	&Exposure times	 \\
\hline	   
5344		&Bailyn$^1$		&WFPC2/PC	&March 1995	&F439W		&40-400s\\
		&			&			&			&F675W		&10-100s\\

*7615		&Sahu$^2$	&WFPC2/WFC	&March-June 1999			&F606W		&260s\\
		&			&			&March 1999 - February 2000	&F814W		&260s\\

*8174		&Van Altena	&WFPC2/WFC	&June 2000	&F555W		&26s\\

10524	&Ferraro		&WFPC2/PC	&April 2006	&F555W		&30s\\

10775	&Sarajedini$^3$	&ACS/WFC	&April 2006	&F606W		&55s\\
		&			&			&			&F814W		&65s\\

*11233	&Piotto$^4$	&WFPC2/WFC	&April 2008	&F450W		&350s\\
		&			&			&			&F814W		&100s\\

11975	&Ferraro		&WFPC2/PC	&April 2009	&F555W		&10s\\

12193	&Lee			&WFC3/UVIS		&May 2011		&F467M		&361-367s\\
		&			&				&				&F547M		&75s\\

12311	&Piotto$^5$		&WFC3/UVIS		&Sept 2010 - March 2011		&F275W		&812s\\
		&			&				&						&F814W		&50s\\

13297	&Piotto$^5$		&WFC3/UVIS		&July 2014		&F336W		&475s\\
		&			&				&				&F438W		&141s\\

\hline
\hline
  \end{tabular}
  \caption{\textit{HST} archival data of M~22 that satisfied our observing criteria. An asterisk in front of the program number denotes WFPC2 data sets which we subsequently excluded (see text). Reference papers describing the observing programs in detail, when available, are: $^1$\cite{bailyn96}; $^2$\cite{sahu01}; $^3$\cite{sarajedini07}; $^4$\cite{piotto12}; $^5$\cite{piotto15}. \label{tab:data}}
  \end{center}
\end{table*}
%% =====================================================

\subsection{Reduction}

We reduced available images from data sets given in \Tab{tab:data} using the suite of tools developed by Jay Anderson \cite[e.g.][]{anderson03}. For ACS and WFC3 images, we used {\tt flc} images, which are available in the archive, and have been corrected for charge-transfer efficiency (CTE) losses. The importance of CTE loss correction has been discussed in several papers; for more details, see e.g. \cite{bellini14}.

The first detection and measurement of stars in each image was made using the routine {\tt hst1pass}. When available, a standard PSF array was used for each instrument/ filter combination \cite{anderson06}; when these were not available, we used the available standard PSF that was closest in wavelength (e.g. for the F547M filter, we used the F467M standard PSF). This is satisfactory because the PSF that is used for the measurement of each star is actually an interpolation of the nearest four PSFs on the image. Furthermore, the time dependence is also accounted for by calculating perturbations to the PSF for each observation. Finally, we also applied the geometric distortion corrections of \cite{anderson06} to position measurements.

Using an image from the ACS Survey of Galactic Globular Clusters (\textit{HST} Treasury Program, \citealt{sarajedini07}) as a reference, we used {\tt xym2mat} to derive positions in a common coordinate system of all stars in each image. This is done by deriving a 6-parameter transformation between an image and the reference, allowing us to measure the position of stars on each transformed image. Applying this transformation to the entire set of available images then yields astrometric time series for each star.

Although transformations were derived for WFPC2/WFC images, these were found to be imprecise, with scatter as large as 1 pixel in most measurements. There are several reasons for this: WFPC2's larger pixels, its smaller dynamic range, and the lack of dithering typical of early HST observations. Due to this, we did not include WFPC2/WFC images in the rest of the analysis. This issue did not affect the planetary camera (PC) on WFPC2, and we retain these data for further analysis, in addition to ACS and WFC3 data. The excluded data sets were those of proposal ID 7615 (PI: Sahu), 8174 (PI: Van Altena), and 11233 (PI: Piotto).

\subsection{Proper motion corrections}

Cluster stars are extremely unlikely to be lensed by another object in the cluster, since for such a case, $\ds/\dl \sim 1$, resulting in a very small Einstein ring radius (Eq. \ref{eq:thetae}). We can therefore assume that cluster stars move in a straight lines, i.e. without any microlensing deflections, and use this to correct for systematics that may result in non-rectilinear time series for those stars.

To do this, we fitted straight-line proper motions to the time series of all cluster stars, and looked at the distribution of residuals from those fits. For each filter/ epoch/ instrument, we then computed the median residual from the straight-line fit. In case systematics offsets are present, these median residuals would be nonzero, whereas for random scatter, the distribution of residuals would be centered on zero. For each filter/ epoch/ instrument, we then subtracted this median offset value from the time series of all stars. We binned data taken with the same instrument in 10-day bins; given the very slow-evolving signals we are looking for in this work, any variation that may happen on 10-day timescales can be ignored without losing significant information on the long-timescale events. Finally, position measurements for different filters were combined at each epoch, producing a single position at each epoch/ instrument combination.

\subsection{Stars near the center}

We used the center coordinates for the cluster center from \cite{goldsbury10}, $\alpha=18.36.23.94, \delta=-23:54:17.10$; associated uncertainties are $\pm 0.8\arcsec$. Although any IMBH present in the cluster is expected to be located at the center, $N-$body simulations suggest that it might also be slightly off-center due to interactions between the IMBH and stars in the cluster \citep{devita18}. We therefore selected stars within a radius of 6$\arcsec$ of the center for further analysis, in order to be conservative in accounting both for uncertainties in the precise location of the center itself, and for the amount by which an IMBH might be off-center \citep{devita18}.

\begin{figure}
  \centering
  \includegraphics[width=7cm, angle=0]{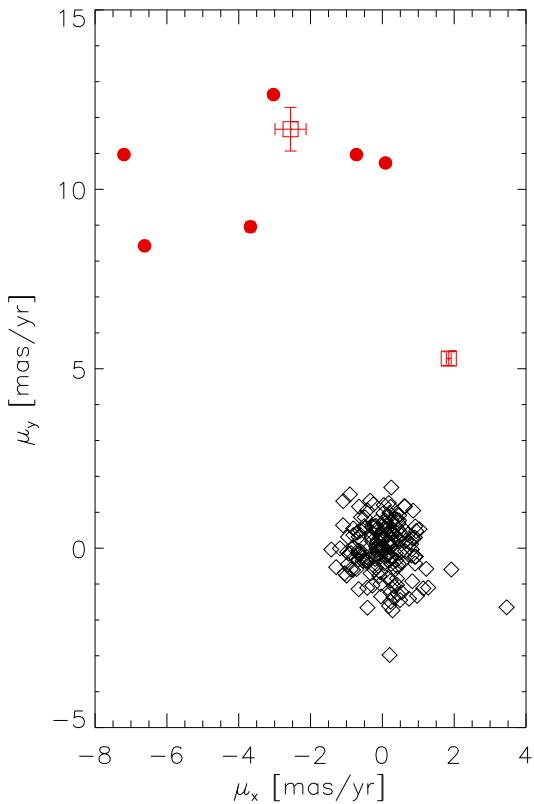}
  \caption{Proper motions derived from initial straight-line fits to astrometric time series. Cluster stars are shown as open diamonds, while Bulge stars, clearly separated in the proper motion diagram, are plotted as filled red circles and open squares (for the two stars with $<4$ epochs with astrometric measurements), with 1-$\sigma$ error bars (too small to see for most stars). \label{fig:allpm}}
\end{figure}

The large M~22-Bulge relative motion of $\sim12.2$ mas yr$^{-1}$ \citep{kains16, bellini14} makes cluster members easily distinguishable from background Bulge stars using their proper motion, derived from initial linear fits to their astrometric time series, as shown in \Fig{fig:allpm}. Using this, we identified 8 Bulge stars from 199 stars detected on the reference image within 6$\arcsec$ of the center. Coordinates and proper motions for these 8 stars are given in \Tab{tab:starlist}

% =====================================================
% =====================================================

%% =====================================================
\begin{table*}
  \vspace{0.5cm}
\begin{center}
  \begin{tabular}{cccccccc}
\hline
Star  	&RA	&Dec	&$\mu_x$	&$\mu_y$								&$N_e$ \\
  	&(J2000.0)	&(J2000.0)	&[mas yr$^{-1}$]	&[mas yr$^{-1}$]		& \\
\hline
1	&18:36:24.19	&-23:54:18.19	&-0.72$\pm$0.02	&10.97$\pm$0.03	&8 \\
2	&18:36:24.43	&-23:54:20.52	&-3.67$\pm$0.12	&8.96$\pm$0.02	&5 \\
3	&18.36:24.35	&-23:54:14.04	&-7.19$\pm$0.02	&10.97$\pm$0.08	&6 \\
4	&18:36:24.24	&-23:54:12.60	&-2.56$\pm$0.43	&11.68$\pm$0.61	&3 \\
5	&18:36:24.22	&-23:54:18.30	&-6.62$\pm$0.04	&8.42$\pm$0.02	&6 \\
6	&18:36:24.13	&-23:54:15.00	&1.85$\pm$0.07	&5.28$\pm$0.19	&4 \\
7	&18:36:23.88	&-23:54:18.27	&-3.03$\pm$0.03	&12.64$\pm$0.05	&7 \\
8	&18:36:23.87	&-23:54:19.19	&0.09$\pm$0.05	&10.73$\pm$0.04	&8 \\

\hline	   

  \end{tabular}
  \caption{Coordinates, proper motions (along the $x$ and $y$ image directions), and number of epochs with astrometric measurements for the 8 Bulge stars in our sample within 6$\arcsec$ of the cluster center. \label{tab:starlist}}
  \end{center}
\end{table*}
%% =====================================================

\section{Modelling procedure}\label{sec:modelling}

For each star in our sample, we fitted two models: 

\begin{enumerate}
\item A proper-motion (PM)-only model, with parameters $\mu_x, \mu_y, x_0$, and $y_0$, respectively the proper motions along the $x$ and $y$ directions on the image, and arbitrary reference $x$ and $y$ positions 
\item An astrometric microlensing (ML) model, described by the standard microlensing parameters $\tz, \te, \uz$, described in \Sec{sec:microlensing}, as well as $\alpha$, the relative angle of the lens-source trajectories, the Einstein ring radius $\thetae$, as well as the proper motion parameters $\mu_x, \mu_y, x_0$, and $y_0$.
\end{enumerate}

Note that the closest star to the cluster center is $\sim 1.6\arcsec$ away from the center coordinates of \cite{goldsbury10}, which means that any microlensing event would either occur in the unresolved regime, or be very well approximated by unresolved microlensing models, as long as the Einstein ring radius is less than $\sim 400$ mas, which, for a lens in M~22 and a Bulge source, corresponds to a $\sim10^5\msun$ IMBH. We also note here that in the following discussion, we use the term ``shift" when referring to microlensing fits, and ``residuals" when referring to residual positions after subtraction of a proper motion model.

After fitting both the PM and ML models, we compared them using the Bayesian Information Criterion (BIC, \citealt{schwarz78}), defined as

\begin{equation}\label{eq:bic}
\mathrm{BIC}=-2\ln(\mathcal{L}) + \np\ln(\nd)\, - \np\ln(2\pi)\, ,
\end{equation}
\noindent
where $\mathcal{L}$ is the likelihood, $\nd$ is the number of data points (twice the number of epochs, with astrometric measurements, since the position is measured in two directions), and $\np$ is the number of model parameters (9 for ML models, 4 for PM models). Here we have assumed constant priors on most parameters, as no information is available from a photometric microlensing light curve fit, for instance \citep[e.g.][]{kains17} . However, we do impose a Gaussian prior on the relative source-lens motion, using the cluster-Bulge relative motion $12.2\pm 3.9$ mas yr$^{-1}$ measured by \cite{kains16} from the proper motion catalogue of \cite{bellini14}. This is particularly important in cases where data are limited, in which case model fits may converge to parameters that are not consistent with the observed relative motion. Using the BIC for the best-fit MCMC models, we then compute their relative probability,

\begin{equation}\label{eq:bicprob}
\frac{P(\mathrm{PM})}{P(\mathrm{ML})}=\exp(0.5(\mathrm{\dbic)})\, ,
\end{equation}
\noindent
where $\dbic=\mathrm{BIC_{ml}}-\mathrm{BIC_{pm}}$. In addition, we impose an additional requirement that an astrometric lensing be at least 1000 times more probable to be favoured, which corresponds to a threshold $\dbic$=13.82. The BIC is usually most appropriate when $\nd >> \np$, but the high additional penalty we impose on the ML models means that even with a relatively small number of data points, the BIC is sufficient for the purposes of selecting a model in this work. \textbf{To test this, we generated PM-only astrometric time-series, with scatter corresponding to our real data, and fitted both PM and ML models to these simulated data sets. We generated and fitted 1000 such data sets, and found that with such a strict penalty on the ML models, the ``false alarm" rate, i.e. the fraction of these events where the ML model is favored over the correct PM model, is negligible ($<0.1\%$).}

\begin{figure}
  \centering
  \vspace{0.3cm}
  \includegraphics[width=8.5cm, angle=0]{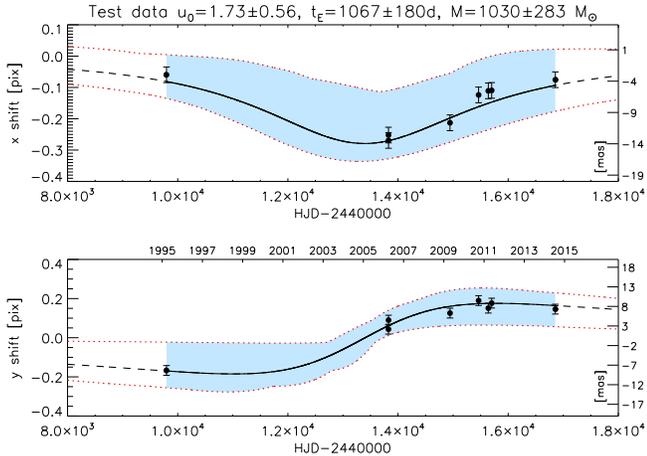}
  \caption{Best-fit model to test data for a $10^3 \msun$ IMBH in M~22, with $\te=1900$ d, $\uz=1.5$. The source-lens relative motion is $\sim14$ mas yr$^{-1}$. Recovered parameters are in agreement with input parameters, but with large error bars. Blue shaded areas and red dotted lines show the limits of the 99.7\% confidence interval. The top axis in the bottom panel gives the time in calendar years. \label{fig:modeltest}}
\end{figure}

Since good astrometry could not be extracted from WFPC2/WFC data, this left us with significant gaps in the time series, and only 8 distinct epochs. To assess whether such coverage would be sufficient to recover lensing parameters, we generated synthetic astrometric time series from a set of input parameters (IMBH mass, source-lens relative motion, microlensing impact parameter). We assumed Gaussian errors on astrometric measurements and scatter consistent with our real data. We then analyzed these synthetic data sets with our modelling codes, and found that synthetic astrometric curves allowed us to recover input parameters for a wide range of masses, although with large associated uncertainties, as expected with this few epochs. \Fig{fig:modeltest} shows an illustration of one such a fit, for a $10^3\msun$ IMBH, with $\te=1900$ d. However, we do find that this time coverage would miss many events caused by low-mass ($\lesssim5\times10^2\msun$) or very high-mass ($\gtrsim10^5\msun$), due to their short timescales (most of the event happening in coverage gaps) or inability to distinguish the total motion from PM-only motion, respectively.

% =====================================================
% =====================================================

\section{Results and discussion}\label{sec:results}

As expected, all cluster stars are found to be consistent with rectilinear, constant
proper-motion-only models. Given the cluster distance of 3.2 kpc, parallax shifts of $\sim 0.3$ mas are present in the astrometry, corresponding to variations of $\sim0.6\%$ of a pixel. This is smaller than the astrometric precision we achieve in this study, and we can therefore safely conclude that parallax does not have a significant effect on our results. 

Out of 8 Bulge stars, 2 had insufficient ($<4$) epochs with astrometric measurements, due to their close proximity to saturated stars in most images. This leaves 6 Bulge stars with enough data (5-8 epochs) to detect an astrometric microlensing signal. The lensing model is not favoured over the proper-motion-only model for any of these 6 Bulge stars. 

\begin{figure}
  \centering
  \includegraphics[width=9cm, angle=0]{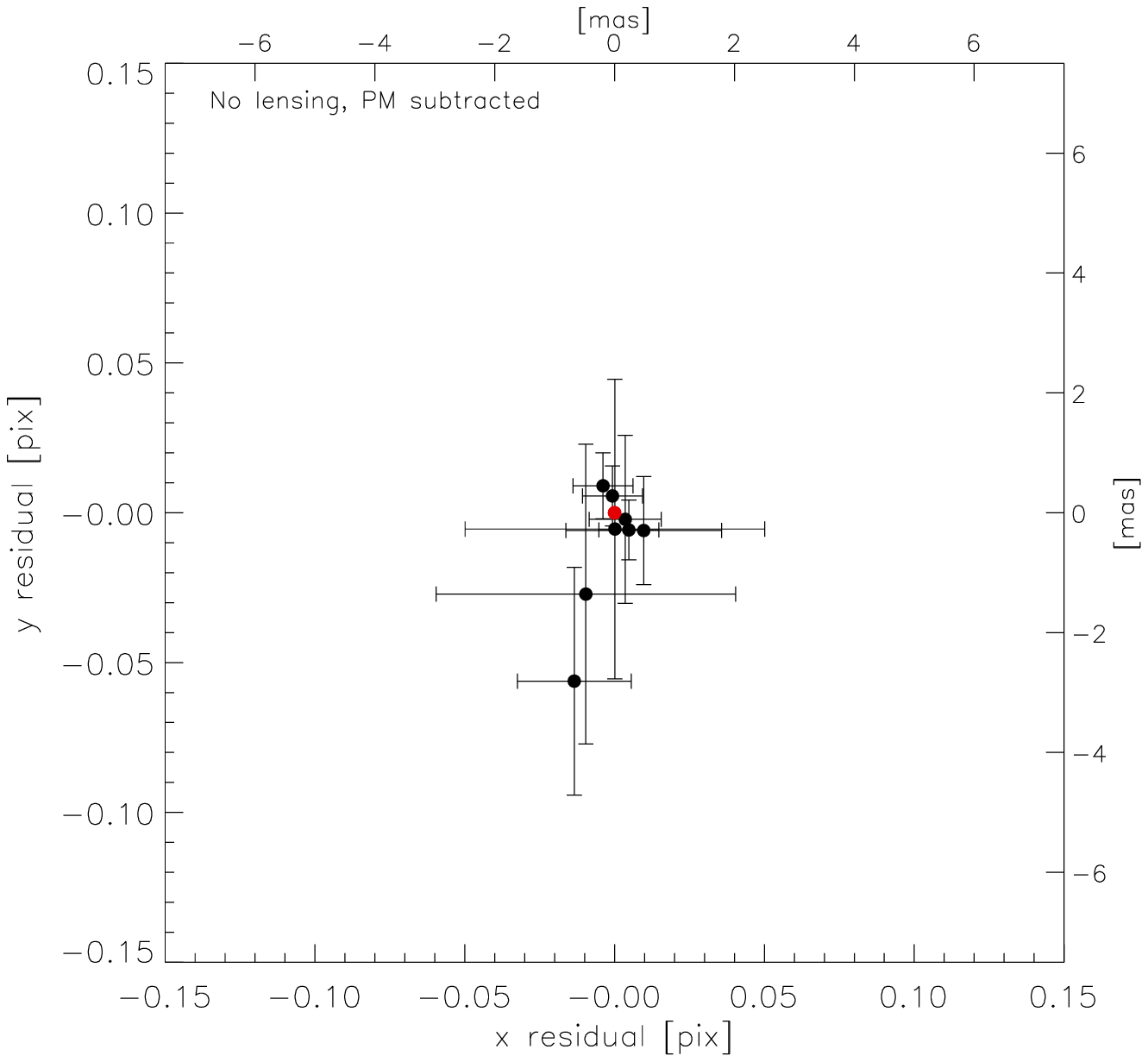}
  \includegraphics[width=9cm, angle=0]{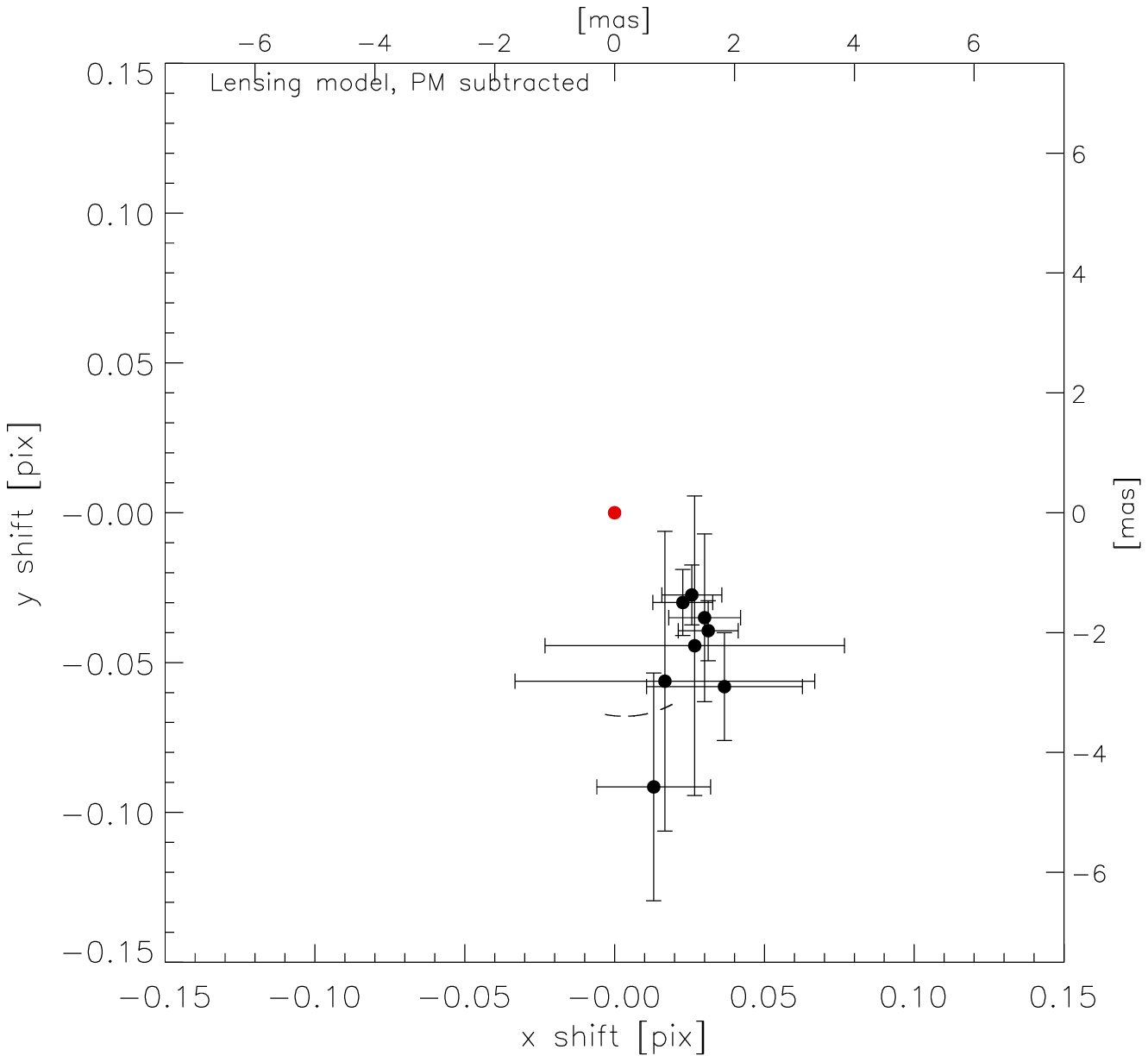}
  \caption{2-D residual positions after proper motion subtraction, for the PM-only model (top) and astrometric microlensing model (bottom), for the closest star to the center of M~22. 
  \label{fig:37299_2d}}
\end{figure}

In \Fig{fig:37299_2d}, we show plots of the time series for the closest star to the center position of \cite{goldsbury10}, in 2-dimensions with the proper motion subtracted, and of the motion in each of the $x$ and $y$ direction, along with the best-fit proper-motion and lensing models.

% =====================================================
% =====================================================

% =====================================================
\begin{figure}
  \centering
  \includegraphics[width=8cm, angle=0]{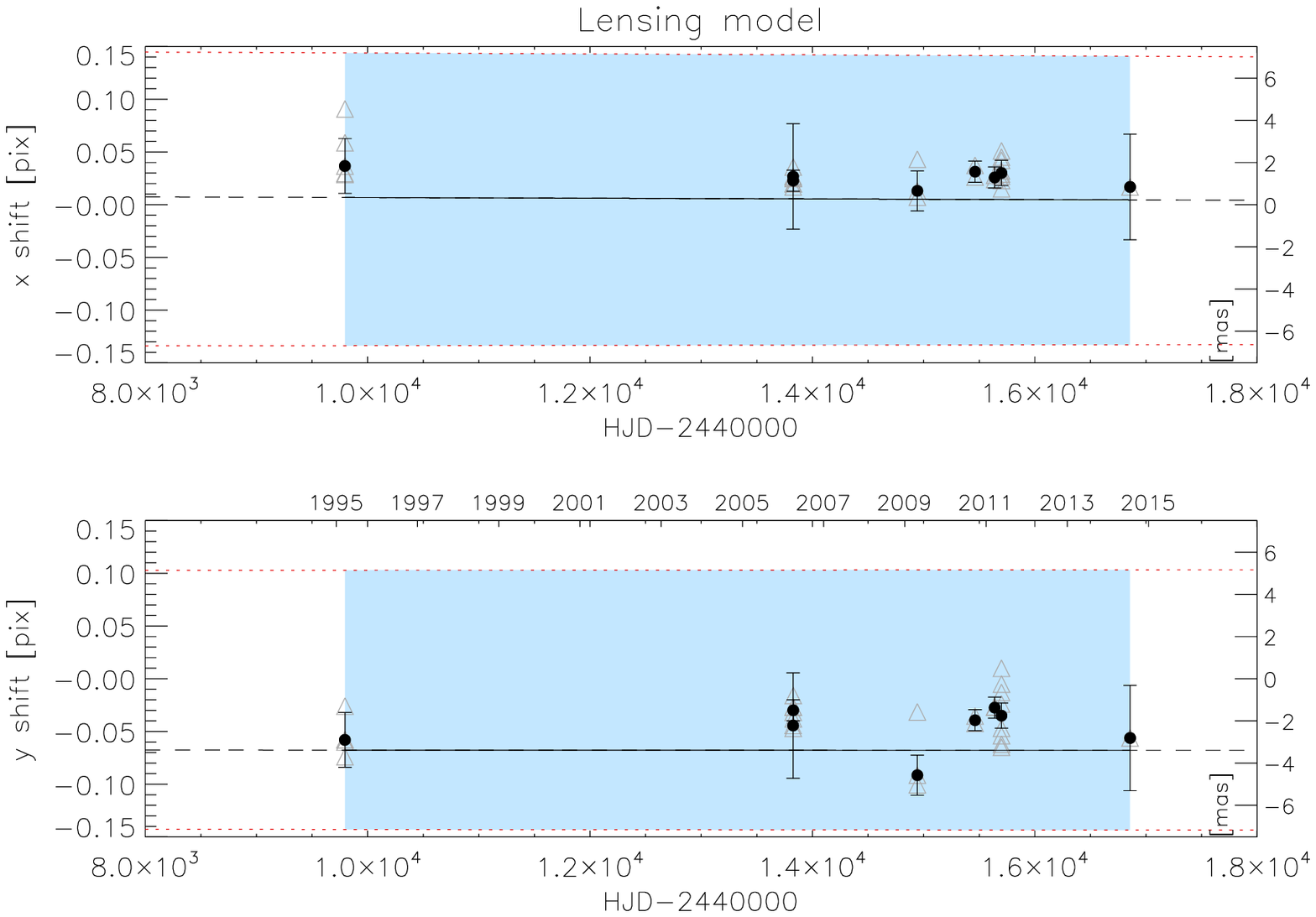}
  \includegraphics[width=8cm, angle=0]{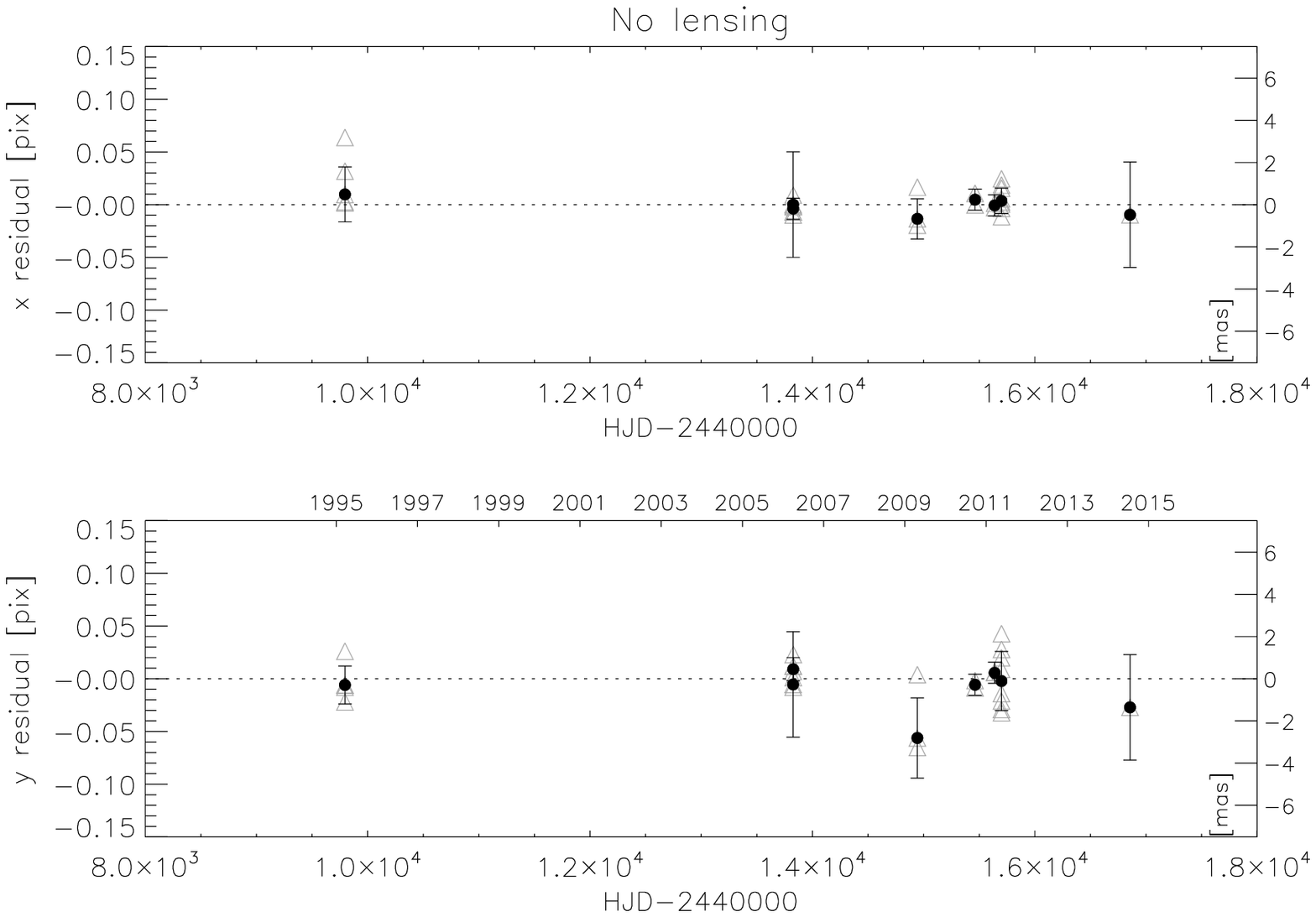}
  \caption{Residual positions after proper motion subtraction, for each of the $x$ and $y$ pixel positions, for the PM-only model (top) and astrometric microlensing model (bottom), for the closest star to the center of M~22. For the microlensing model, the best-fit model to the microlensing shift over the 19.33 year baseline is also plotted as a solid line, and blue shaded areas and red dotted lines show the limits of the 99.7\% confidence interval. For each model, a time axis in calendar years is shown between the two panels. Individual-image measurements are shown as grey open triangles, while combined measurements are plotted as filled circles with error bars.
  \label{fig:37299_1d}}
\end{figure}
% =====================================================

From Figs. \ref{fig:37299_2d}-\ref{fig:37299_1d}, it is clear that the data is consistent with no microlensing deflection taking place. The PM-only model is strongly favoured, and the microlensing fits converge toward very slow-evolving models, which are essentially indistinguishable from PM-only models. These plots are representative of all 6 Bulge stars in our sample.

Although the 10-year gap in our data coverage does not allow us to place strong limits, can we exclude part of the IMBH mass range, based on the lack of detection of an astrometric lensing signal in those 6 stars? From \cite{mclaughlin05}, the larger estimate for the total mass of M~22 is $\log_{10}(M_{\mathrm{tot}}/\msun)\approx 5.64\pm0.05$. Therefore an extreme scenario in which an IMBH makes up $\sim 10\%$ of the total cluster mass would mean an IMBH with $\log(M_{\mathrm{tot}}/\msun)\approx 4.64$. How far down the IMBH mass range would a black hole (BH) have caused a detectable event, given our data?

To estimate this, we carried out Monte Carlo simulations, first placing a BH at a location near the center, with variation in the BH location given due to the uncertainty in the precise location of the center \citep{goldsbury10}, as well as potential IMBH "wander", the size of which goes as $M_{\mathrm{BH}}^{-0.44}$ \citep{devita18}. Given a BH position, the actual location of a Bulge star at observed times, as well as the star's observed motion relative to the cluster (and thus BH), we then derived lensing parameters. Using these, we calculated the expected astrometric deflection at each observed epoch, and determined whether such a deflection would have been detected, given our astrometric measurements.

\begin{figure}
  \centering
  \includegraphics[width=8cm, angle=0]{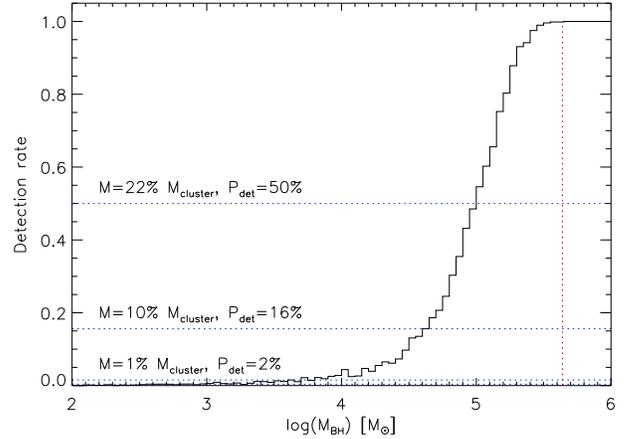}
  \caption{Detection rates as a function of BH mass, given the actual astrometric time-series data, for the star closest to the center of M~22. The vertical dotted red line shows the total cluster mass from \cite{mclaughlin05}, while horizontal lines show detection rates of 50\%, as well as for masses corresponding to 10\% and 1\% of the total cluster mass. Note that an event being detected does not mean that the lens mass can be recovered; see \Fig{fig:limdet} for an estimate of the probability of events being detected \textit{and} have the BH mass recovered. \label{fig:37299_limits}}
\end{figure}

If so, our data allows us to exclude the presence of such a BH at the center of M~22. This was repeated for $10^5$ random BH locations and masses, after which we calculated the detection rate as a function of BH mass, giving us an estimate of the maximum BH mass that would have produced a given detection rate. This is plotted in \Fig{fig:37299_limits} for the closest star to the cluster center. For that star, the detection rate rises steadily as a function of mass, due to the small impact parameter, meaning that high-mass BH events would have been detected despite unfolding slowly. A 50\% detection rate corresponds to a BH mass of $\sim 22\%$ of the total cluster mass, while for a BH mass of 10\% and 1\% of the cluster mass, the detection rates would be $\sim$16\% and 2\%, respectively. Finally, we also estimated the detection rates, adding a data point every two years over our coverage gap (1996-2005), with an error bar corresponding to the median error in our data set, and Gaussian scatter around a straight-line source trajectory, to assess the improvement from having better time coverage. For the simulated data sets, we find a 50\% detection rate at M$\sim$0.12$M_{\rm cluster}$, as well as 42\% and 2\% for BH masses of 10\% and 1\% of the total cluster mass, respectively. As expected, more regular time coverage improves detection rates significantly for higher masses ($\gtrsim 10^4\msun$), whereas more frequent observations are needed to improve detection rates at lower masses ($\lesssim 10^4\msun$).

% =====================================================
% =====================================================

\begin{figure}
  \centering
  \includegraphics[width=8cm, angle=0]{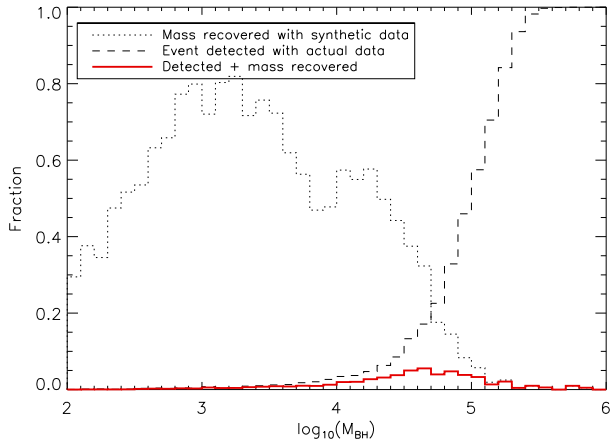}
  \caption{Mass recovery rate (dotted line), detection rate (dashed line), and the product of these two, giving the rate of events that are detected and have their BH mass recovered, as a function of BH mass.\label{fig:limdet}}
\end{figure}

In addition to detecting an event caused by the presence of the BH in the cluster, how often can we also recover the mass of the BH, given our time sampling? We can use simulated data (e.g. Fig. \ref{fig:modeltest}) to estimate how the mass recovery rate changes with BH mass, assuming the time coverage of our data set. We did this by generating synthetic data sets, fitting these with our astrometric modelling code, and comparing the fitted parameters to the input parameters. In this case, we considered the mass recovered when the fitted mass was within 0.5 dex of the input mass. In \Fig{fig:limdet}, we plot the mass recovery rate, detection rate, and the multiplication of these two, giving the rate of events that are detected \textit{and} for which the mass is recovered. We find that given our data, only BH with masses between $10^4$ and $10^5\msun$ have a significant chance of both being detected and have their mass recovered, between $\sim$ 3 and 6\%. This is because higher-mass events, while they can be detected because of the large signal they produce, are slow-evolving, leading to mostly rectilinear deflections on timescales of $\sim$years; these do not provide good constraints on the lens masses. On the other hand, lower-mass events can be better modeled thanks to less rectilinear deflection signals, but are less easily detected because of the smaller signals they produce.

% =====================================================
% =====================================================

Finally, we checked the catalogue from the \textit{Gaia} \citep{gaia16} data release 2 (DR2, \citealt{gaiadr2}) for matches to the Bulge stars in our sample, to compare our derived proper motions with DR2 values. Indeed, in the most extreme cases involving a background star being deflected by a very massive BH, with a low impact parameter, a trajectory might appear rectilinear over even a 20-year baseline, but could actually be a small fraction of a lensing event. In such a case, the true direction of the star's proper motion would be different from that derived from the observations, and a second set of later observations might reveal a change in direction, hinting at possible microlensing taking place. Although this would be unlikely to be seen from \textit{Gaia} data, since they only span 22 months of data from July 2014 to May 2016, whereas our archival data set ends in 2014, we considered this possibility for thoroughness. Unfortunately, only three of our Bulge sources had \textit{Gaia} matches at a distance $<1\arcsec$, and none of these 3 had proper motion values in DR2. However, this is a check worth making when considering the possibility of astrometric microlensing by very massive BHs.

\section{Conclusions}\label{sec:conclusion}

Using \textit{HST} archival data spanning nearly 20 years of baseline, we derived astrometric time series for 199 stars within 6$\arcsec$ of the center of M~22, but with a large 10-year gap, due to being unable to extract high-precision astrometry from the WFC chips of the WFPC2 instrument. This includes 6 Bulge stars with sufficient time coverage to detect microlensing signals, if present, and constrain the lens properties. These 6 stars do not show any signatures of microlensing deflections, and are consistent with proper-motion-only models. Although in theory it is possible to use these non-detections to place limits on the mass of any IMBH present at the cluster center, the insufficient precision of WFPC2/ WFC astrometry led to a large gap in time coverage that meant we were unable to derive any useful limits for realistic BH masses.

We conclude that it is worth extending this work to other clusters, particularly those with rich \textit{HST} archival data sets, such as M~4 or 47 Tuc. The tools developed in this study will make future analyses of other clusters much faster. In addition, the astrometric microlensing analysis tools we wrote as part of this project would also be useful in preparation for the Wide Field Infrared Survey Telescope (\textit{WFIRST})'s Microlensing Survey, for which astrometry has the potential to significantly improve the constraints yielded by the survey on lens masses \citep{gould14}, including on the mass function of stellar-mass BHs and extrasolar planets. In addition, it will be worth revisiting models in the future, for instance after observations of the centers of these clusters have been taken by the \textit{James Webb Space Telescope (JWST)}, which will afford even better astrometric precision than what is currently achievable with \textit{HST}.

\section*{Acknowledgements}

Support for this work was provided by NASA through grant number AR-14571 (PI: Kains) from the Space Telescope Science Institute, which is operated by AURA, Inc., under NASA contract NAS 5-26555. Based on observations made with the NASA/ESA Hubble Space Telescope, obtained by the Space Telescope Science Institute. All of the data presented in this paper were obtained from the Mikulski Archive for Space Telescopes (MAST). 

%\software{\texttt{hst2xym} \citep{anderson06}

\bibliographystyle{aasjournal}
\bibliography{thesisbib}

\label{lastpage}

\end{document}